\documentclass[12pt]{article}
\usepackage{graphicx}

\newcommand{\ket}[1]{|#1\rangle}

\newtheorem{proposition}{Proposition}
\newtheorem{theorem}[proposition]{Theorem}
\newtheorem{problem}[proposition]{Problem}

\newtheorem{corollary}[proposition]{Corollary}

\newenvironment{proof}{\noindent\emph{Proof.}\ }{\hfill\rule{1.3ex}{1.3ex}}

\begin{document}
\title{Algebraic geometric construction of a quantum stabilizer code}
\author{Ryutaroh Matsumoto\\
Department of Communications and Integrated Systems\\
Tokyo Institute of Technology, 152-8552 Japan\\
Email: \texttt{ryutaroh@rmatsumoto.org}}
\date{August 8, 2001}
\maketitle
\begin{abstract}
\noindent
The stabilizer code is the most general algebraic construction of quantum
error-correcting codes proposed so far.
A stabilizer code can be constructed from a self-orthogonal
subspace of a symplectic space over a finite field.
We propose a construction method of such a self-orthogonal
space
using an algebraic curve.
By using the proposed method we construct an asymptotically good
sequence of binary stabilizer codes.
As a byproduct we improve the Ashikhmin-Litsyn-Tsfasman bound
of quantum codes.
The main results in this paper can be understood without
knowledge of quantum mechanics.
\end{abstract}

\section{Introduction}
Recently quantum computation
and quantum communication have attracted much
attention,
because the use of quantum mechanical phenomena
can offer
unusual efficiency in computation and
communication.
We have to protect quantum states from environmental noise
in quantum computation and some methods in quantum communication,
such as the quantum superdense coding \cite{bennett01,bennett92}.
The quantum error-correcting codes (or quantum codes) independently
proposed by Shor \cite{shor95} and Steane \cite{steane96b}
is one of techniques for protecting quantum states.
Recently it was recognized that construction of quantum codes
is connected with the problem of finding a linear space over
a finite field with certain properties (see Theorem \ref{th1}).
In this paper we propose a method of constructing
such linear spaces from an algebraic curve.

Let us explain quantum codes and their connection with
linear spaces over finite fields.
We begin with the notion of $t$-error correction.
Let $\mathcal{H}$ be a $q$-dimensional complex linear space,
where $q$ is a prime power,
and suppose that $\mathcal{H}$ represents
a physical system of interest.
A quantum code $Q$ is a $q^k$-dimensional
subspace of $\mathcal{H}^{\otimes n}$.
When we want to protect a quantum state in $\ket{\varphi}\in
\mathcal{H}^{\otimes k}$,
we encode $\ket{\varphi}$ into a state in $Q$.
So we encode a quantum state of $k$ particles
into that of $n$ particles.
Suppose that we send $\ket{\varphi}\in Q$
and receive $\ket{\psi}\in \mathcal{H}^{\otimes n}$.
A quantum code $Q$ is said to be
\emph{$t$-error-correcting}
if we can decode $\ket{\varphi}$ from $\ket{\psi}$
provided that at least the states of $n-t$ particles
in $\ket{\psi}$ are left unchanged from $\ket{\varphi}$.

Since a change of a quantum state is continuous,
the notion of $t$-error correction seems
nonsense at first glance \cite{kak99}.
This notion can be justified as follows:
In general the decoding process of a quantum code
does not decode perfectly the transmitted quantum state
from a received one.
However,
the decoded state and a transmitted state
become closer as $t$ increases provided that
the quantum channel used is memoryless as a $q$-ary channel
\cite[Section 7.4]{preskill98},
\cite{matsumotoerror}.
A quantitative relation between the closeness of states,
the noisiness of a channel, and $t$ can be found in \cite{matsumotoerror}.

In \cite{matsumotoerror}
it is shown that one can make 
the decoded state arbitrary close to the transmitted state 
by increasing the code length provided that
the ratio $t/n$ is fixed and is sufficiently large compared with
the noisiness of the channel.
This is a major motivation for studying long codes
as in the classical coding theory \cite[Section 4.3]{peterson}.
When the code length is small,
we can find good quantum codes by examining all possible codes
by a computer.
However,
when the code length is large,
we need some systematic construction method
for producing good quantum codes.
In this paper we propose such a method.

We are now able to state the connection between
quantum codes and finite fields,
which is obtained for the binary code ($\dim \mathcal{H} = 2$) by
Calderbank et~al. \cite{calderbank97,calderbank98} and
generalized to the nonbinary case by Ashikhmin and Knill
\cite{ashikhmin00}.
Let $\mathbf{F}_q$ be the finite field with $q$ elements.
For vectors $\vec{x} = (x_1$, \ldots, $x_{2n})$ and
$\vec{y} = (y_1$, \ldots, $y_{2n})\in \mathbf{F}_q^{2n}$
we define the standard symplectic form (or alternating form) by
\begin{equation}
\langle \vec{x},\vec{y}\rangle_\mathrm{s}
= \sum_{i=1}^n x_i y_{n+i} - \sum_{i=1}^n x_{n+i} y_{i}.
\label{symplectic}
\end{equation}
A linear space with a nondegenerate symplectic form
is called a symplectic space.
For a vector $\vec{x}\in \mathbf{F}_q^{2n}$
define the weight of $\vec{x}$ by
\begin{equation}
w(\vec{x}) = \sharp\{ 1\leq i \leq n \mid (x_i,x_{n+i}) \neq (0,0)\}.
\label{eq:weight}
\end{equation}
For a subspace $C \subset \mathbf{F}_q^{2n}$
we define
\[
C^{\perp\mathrm{s}} = \{ \vec{x}\in\mathbf{F}_q^{2n} \mid
\forall \vec{y}\in C,\;
\langle \vec{x},\vec{y}\rangle_\mathrm{s} = 0\},
\]
that is, the orthogonal space of $C$ with respect to (\ref{symplectic}).

\begin{theorem}\label{th1}\cite{ashikhmin00,calderbank97,calderbank98}
If there is an $(n+k)$-dimensional subspace $C \subset \mathbf{F}_q^{2n}$
such that $C \supseteq C^{\perp\mathrm{s}}$,
then we can construct a $\lfloor (d(C\setminus C^{\perp\mathrm{s}})-1)/2\rfloor$-error-correcting quantum code $Q \subset \mathcal{H}^{\otimes n}$ of
dimension $q^k$,
where
\[
d(C\setminus C^{\perp\mathrm{s}}) = \min\{w(\vec{x}) \mid
\vec{x}\in C\setminus C^{\perp\mathrm{s}}\}.
\]
\end{theorem}

The quantum code $Q$ constructed by this method
is called a \emph{stabilizer code} and
is proposed independently by Gottesman \cite{gottesman96}
and Calderbank et~al.\ \cite{calderbank97,calderbank98}.
The nonbinary generalization is due to
Knill \cite{knill96a} and Rains \cite{rains97}.
The stabilizer code is the most general algebraic construction
of quantum codes proposed so far.

The value $d(C\setminus C^{\perp\mathrm{s}})$ is called
the \emph{minimum distance} of a stabilizer code.
A stabilizer code with minimum distance
$d$ encoding $k$ particles into $n$ particles
is called an $[[n,k,d]]$ code.

Rains \cite[p.1831, Remarks]{rains97}
observed that a $q_1 q_2$-ary stabilizer code
is a tensor product of a $q_1$-ary stabilizer code
and a $q_2$-ary one
if $q_1$ and $q_2$ are relatively prime.
So we restrict ourselves to stabilizer codes
for quantum systems of prime power dimension.

This paper is organized as follows:
In Section \ref{sec2}
we propose a construction method of quantum stabilizer codes
from algebraic curves and discuss decoding process of 
the constructed codes.
In Section \ref{sec31}
we construct an asymptotically good sequence of quantum codes
as an example of the proposed construction method.
In Section \ref{sec32}
we improved the construction of asymptotically good
sequence in \cite{ashikhmin01}
as a byproduct of the construction in Section \ref{sec31},
and compare the sequences in Section \ref{sec3}
 with the known asymptotically good sequences
\cite{ashikhmin01,chen01b} in Figure \ref{fig1}.

\sloppy

\section{Quantum stabilizer codes from
algebraic curves}\label{sec2}
\subsection{Construction}
We shall use the formalism of algebraic function fields
instead of algebraic curves.
Notations used are exactly the same as those in
Stichtenoth's textbook \cite{bn:stichtenoth}.

\fussy

\begin{proposition}\label{prop2}
Let $F/\mathbf{F}_q$ be an algebraic function field of
one variable,
$\sigma$ an automorphism of order $2$
of $F$ not moving elements in $\mathbf{F}_q$, and
$P_1$, \ldots, $P_n$ pairwise distinct places of degree one
such that $\sigma P_i \neq P_j$ for all $i$, $j = 1$, \ldots, $n$.
Let us introduce a condition on a differential $\eta$:
\begin{equation}
\left\{
\begin{array}{l}
v_{P_i}(\eta) = v_{\sigma P_i}(\eta) = -1,\\
\mathrm{res}_{P_i}(\eta) = 1,\\
\mathrm{res}_{\sigma P_i}(\eta) = -1.
\end{array}\right. \label{etacondition}
\end{equation}
The existence of such $\eta$ is guaranteed by
the strong approximation theorem of discrete valuations
\cite[Theorem I.6.4]{bn:stichtenoth}.
Further assume that we have a divisor $G$ such that $\sigma G = G$,
$v_{P_i}(G) = v_{\sigma P_i}(G) = 0$.
Define
\[
C(G) =\{ (f(P_1), \ldots, f(P_n), f(\sigma P_1), \ldots, f(\sigma P_n))
\mid f\in \mathcal{L}(G)\} \subseteq \mathbf{F}_q^{2n}.
\]
Let
\[
H = (P_1 + \cdots + P_n + \sigma P_1 + \cdots + \sigma P_n)
-G + (\eta),
\]
where $\eta$ is as Eq.\ (\ref{etacondition}).
Then we have $C(G)^{\perp\mathrm{s}} = C(H)$.
\end{proposition}

\begin{proof}
In the following argument,
$\vec{x} = (x_1$, \ldots, $x_{2n})$ and
$\vec{y} = (y_1$, \ldots, $y_{2n})$.
By Proposition VII.3.3 in \cite{bn:stichtenoth} and
the assumption on $\sigma$ and $G$,
we have
\begin{eqnarray}
&&(x_1, \ldots, x_{2n}) \in C(G) \nonumber\\
&\Longleftrightarrow& (x_{n+1}, \ldots, x_{2n}, x_1, \ldots, x_n) \in C(G).
\label{codeautomorphism}
\end{eqnarray}
The assertion is proved as follows:
\begin{eqnarray*}
&& \vec{x} \in C(H)\\
&\Longleftrightarrow& \forall\vec{y}\in C(G),\;
\sum_{i=1}^n x_i y_i - \sum_{i=n+1}^{2n} x_i y_i =0\mbox{ (by Corollary 2.7 of
\cite{stichtenoth88})}\\
&\Longleftrightarrow& \forall\vec{y}\in C(G),\;
\sum_{i=1}^n x_i y_{n+i} - \sum_{i=1}^{n} x_{n+i} y_i=0
\mbox{ (by Eq.\ (\ref{codeautomorphism}))}\\
&\Longleftrightarrow& \vec{x} \in C(G)^{\perp\mathrm{s}}.
\end{eqnarray*}
\end{proof}

\begin{corollary}\label{coro3}
Notations as in Proposition \ref{prop2}.
Assume further that $G \geq H$.
Then we can construct an $[[n,k,d]]$ quantum code $Q$,
where
\begin{equation}
k = \dim G - \dim (G-P_1 - \cdots - P_n - \sigma P_1 - \cdots -
\sigma P_n) - n. \label{eqk}
\end{equation}
For the minimum distance $d$ of $Q$, we have
\begin{equation}
d \geq  n- \left\lfloor\frac{\deg G}{2}\right\rfloor. \label{eqd}
\end{equation}
\end{corollary}

\begin{proof}
Theorem \ref{th1} and Proposition \ref{prop2}
show that we can construct a quantum code from $C(G)$
because $C(G) \supseteq C(G)^{\perp\mathrm{s}} = C(H)$.
By Theorem \ref{th1} we have
$k = \dim C(G)-n$.
Theorem II.2.2 of \cite{bn:stichtenoth}
asserts
\[
\dim C(G) = \dim G - \dim (G-P_1 - \cdots - P_n - \sigma P_1 - \cdots -
\sigma P_n),
\]
which shows Eq.\ (\ref{eqk}).

We shall prove Eq.\ (\ref{eqd}).
Suppose that
$w(f(P_1)$, \ldots, $f(\sigma P_n)) = \delta \neq 0$ for
$f \in \mathcal{L}(G)$.
Then there exists a set $\{i_1$, \ldots, $i_{n-\delta}\}$
such that
$f(P_{i_1}) = f(\sigma P_{i_1}) = \cdots =
f(P_{i_{n-\delta}}) = f(\sigma P_{i_{n-\delta}})=0$, which implies
$f \in \mathcal{L}(G - \sum_{j=1}^{n-\delta} (P_{i_j}+\sigma P_{i_j}))$.
Since $f \neq 0$, we have
\begin{eqnarray*}
&& \dim \left(G - \sum_{j=1}^{n-\delta} (P_{i_j}+\sigma P_{i_j})\right) > 0\\
&\Longrightarrow& \deg \left(G - \sum_{j=1}^{n-\delta} (P_{i_j}+\sigma P_{i_j})\right) \geq 0\\
&\Longleftrightarrow& \deg G - 2(n-\delta) \geq 0\\
&\Longleftrightarrow& 2\delta \geq 2n - \deg G\\
&\Longleftrightarrow& \delta\geq n-\left\lfloor \frac{\deg G}{2}\right\rfloor.
\end{eqnarray*}
\end{proof}

The above construction provides good codes only when $q$ is large
as the classical algebraic geometry codes.
For small $q$, we construct a $q$-ary quantum code from
a $q^m$-ary one by

\begin{theorem}[Ashikhmin and Knill \cite{ashikhmin00}]\label{th:ak}
Let $m$ be a positive integer,
$\{\alpha_1$, \ldots, $\alpha_m\}$ an $\mathbf{F}_q$-basis
of $\mathbf{F}_{q^m}$.
Define $\mathbf{F}_q$-linear maps
$\alpha : \mathbf{F}_q^m \rightarrow \mathbf{F}_{q^m}$ sending
$(x_1$, \ldots, $x_m)$ to $x_1 \alpha_1 + \cdots + x_m \alpha_m$,
and $\beta : \mathbf{F}_q^m \rightarrow \mathbf{F}_{q^m}$ sending
$(x_1$, \ldots, $x_m)$ to 
\[
(\alpha_1, \ldots, \alpha_m) M \left(
\begin{array}{c}
x_1\\
\vdots\\
x_m\end{array}\right) \in \mathbf{F}_{q^m},
\]
where $M$ is an $m \times m$ matrix defined by
$M_{ij} = \mathrm{Tr}_q^{q^m} (\alpha_i \alpha_j)$
with the trace function $\mathrm{Tr}_q^{q^m}$ from
$\mathbf{F}_{q^m}$ to $\mathbf{F}_q$.
For $C \subseteq \mathbf{F}_{q^m}^n$,
let $\gamma(C) = \{ (\alpha^{-1}(x_1)$, \ldots, $\alpha^{-1}(x_n)$,
$\beta^{-1}(x_{n+1})$, \ldots, $\beta^{-1}(x_{2n})) \mid
(x_1$, \ldots, $x_{2n}) \in C\} \subseteq \mathbf{F}_q^{2mn}$.
If $C^{\perp\mathrm{s}} \subseteq C$ for $C \subseteq \mathbf{F}_{q^m}^{2n}$,
then $(\gamma(C))^{\perp\mathrm{s}} \subseteq \gamma(C)$.
We also have $d(\gamma(C)\setminus (\gamma(C))^{\perp\mathrm{s}})
\geq d(C\setminus C^{\perp\mathrm{s}})$.
\end{theorem}

\subsection{Determination of the error operator from
measurement outcomes}
The decoding process of a quantum stabilizer code is usually
proceeded as follows \cite{calderbank98}:
One measures each observable corresponding to
a generator of the stabilizer group of the code,
then determines which unitary operator on $\mathcal{H}^{\otimes n}$ 
should be
applied to the received state.

In the determination of the operator from measurement outcomes
we have to solve the following problem.

\begin{problem}\label{decpro}
Let $C$ be an $(n+k)$-dimensional subspace of $\mathbf{F}_q^{2n}$
such that $C^{\perp\mathrm{s}} \subseteq C$.
Given $s_1$, \ldots, $s_{n-k} \in \mathbf{F}_q$,
find a vector $\vec{e}$ having the minimum weight (\ref{eq:weight})
in the set
$\{ \vec{y} \in \mathbf{F}_q^{2n} \mid
\langle \vec{y}, \vec{b}_i \rangle_\mathrm{s} = 
s_i$ for $i=1$, \ldots, $n-k\}$,
where $\{\vec{b}_1$, \ldots, $\vec{b}_{n-k}\}$ is
a basis of $C^{\perp\mathrm{s}}$.
\end{problem}
If there exists a vector $\vec{e} \in \mathbf{F}_q^{2n}$
such that $\langle \vec{e}, \vec{b}_i\rangle_\mathrm{s} = s_i$
for $i=1$, \ldots, $n-k$ and that
\begin{equation}
2w(\vec{e}) + 1 \leq n - \left\lfloor \frac{\deg G}{2}\right\rfloor,\label{eq:weighte0}
\end{equation}
where $\{\vec{b}_1$, \ldots, $\vec{b}_{n-k}\}$ is a basis of
$C(G)^{\perp\mathrm{s}} = C(H)$ constructed in Corollary \ref{coro3},
then we can efficiently find $\vec{e}$ from
$s_1$, \ldots, $s_{n-k}$ as follows.

The algorithm of Farr\'an \cite{farran00} efficiently
finds the unique vector $\vec{x}$ having the minimum \emph{Hamming}
weight $w_\mathrm{H}(\vec{x})$
in the set 
$\{ \vec{y} \in \mathbf{F}_q^{2n} \mid
\langle \vec{y}, \vec{b}_i \rangle = 
s_i$ for $i=1$, \ldots, $n-k\}$
from given $s_1$, \ldots, $s_{n-k}$,
provided that $2 w_\mathrm{H}(\vec{x}) + 1 \leq 2n - \deg G$,
where $\langle \vec{x}, \vec{b}_i\rangle$ is the standard
inner product of $\vec{x}$ and $\vec{b}_i$
and $\{\vec{b}_1$, \ldots, $\vec{b}_{n-k}\}$ is
a basis of $C(H)$.

Let $\vec{e} = (e_1$, \ldots, $e_{2n})$
and $\vec{e'} = (-e_{n+1}$, \ldots, $-e_{2n}$, $e_1$, \ldots, $e_n)$.
Then $s_i = \langle \vec{e'}, \vec{b}_i\rangle =
\langle
\vec{e},\vec{b}_i\rangle_\mathrm{s}$.
Since 
$w_\mathrm{H}(\vec{e'}) \leq 2w(\vec{e})$, Eq.\ (\ref{eq:weighte0})
implies
\[
2 w_\mathrm{H}(\vec{e'}) + 1 \leq 2n - \deg G,
\]
and the algorithm of Farr\'an finds $\vec{e'}$ from $s_1$,
\ldots, $s_{n-k}$ correctly.
We can easily find $\vec{e}$ from $\vec{e'}$.

\section{Asymptotically good sequence of quantum codes}\label{sec3}
\subsection{Sequence of codes by the proposed method}\label{sec31}
In this section we construct an asymptotically good
sequence of binary quantum codes from
the Garcia-Stichtenoth function field \cite{garcia95}.
Let
$F_i = \mathbf{F}_{q^2}(x_1$, $z_2$, \ldots, $z_i)$ with
\begin{eqnarray*}
z_i^q + z_i - x_{i-1}^{q+1} &=& 0,\\
x_i &=& z_i/x_{i-1}.
\end{eqnarray*}

\begin{proposition}\label{prop:sequence}
For an integer $m \geq 2$
there exists a sequence of $[[n_i,k_i,d_i]]$
binary quantum stabilizer codes such that
\begin{eqnarray*}
\lim_{i\rightarrow\infty} n_i &=& \infty,\\
\liminf_{i\rightarrow\infty} k_i / n_i &\geq& R_m^{(1)}(\delta),\\
\liminf_{i\rightarrow\infty} d_i / n_i &\geq& \delta,
\end{eqnarray*}
where
\[
R_m^{(1)}(\delta) = 1- \frac{2}{2^m-1} - 4m \delta.
\]
\end{proposition}

\begin{proof}
We shall consider the Garcia-Stichtenoth function field $F_i$
over $\mathbf{F}_{2^{2m}}$ with $i\geq 2$. Let $q=2^m$.
Since the Galois group of $F_i/F_{i-1}$ is
isomorphic to the additive group of $\mathbf{F}_2^m$
\cite[Proposition 1.1 (i)]{garcia95},
there exists an automorphism $\sigma \in \mathrm{Gal}(F_i/F_{i-1})$
of order $2$.

Let $n_i = (q^2-1)q^{i-1}/2$, and $y = x_1^{q^2-1}-1$.
The zero divisors of $y$ consist of $2n_i$ places of degree one
\cite[Section 3]{garcia95}.
Let $F_i^\sigma$ be the fixed field of $\sigma$.
Let $Q$ be a zero of $y$.
There exists a zero $Q'$ of $y$ such that $Q' \neq Q$ and
$Q \cap F_i^\sigma = Q' \cap F_i^\sigma$.
Since $F_i/F_i^\sigma$ is Galois,
by Theorem III.7.1 of \cite{bn:stichtenoth} we have $\sigma Q = Q'$.
Therefore we can write the zero divisor of $y$ as
$P_1 + \sigma P_1 + \cdots + P_{n_i} + \sigma P_{n_i}$
such that $\sigma P_j \neq P_l$ for all $j,l$.
Let $\eta = dy /y = x_1^{q^2-2} dx_1 / y$.
By Proposition VII.1.2 of \cite{bn:stichtenoth},
$\eta$ satisfies the condition (\ref{etacondition}).

Let $G'_0 =
(\eta) + P_1 + \sigma P_1 + \cdots + P_{n_i} + \sigma P_{n_i}$,
and $P_\infty$ the unique pole of $x_1$ in $F_i$.
We have
\[
G'_0 = (q^2-2) (x_1) - (q^2-1) v_{P_\infty}(x_1) P_\infty + (dx_1).
\]
The different exponent of $F_i/F_1$ is even at every place of $F_i$
(see the text below Lemma 2.9 of \cite{garcia95}).
Hence the discrete valuation of $(dx_1)$ is even at every place of
$F_i$ by Remark IV.3.7 of \cite{bn:stichtenoth}.
Observe that $v_{P_\infty}(x_1) = -q^{i-1}$ \cite{garcia95}.
Therefore the valuation of
the divisor $G'_0$
is an even integer at every place of $F_i$.
Define $G_0 = G'_0/2$.
We have
\begin{eqnarray*}
\deg G_0 &=& \frac{2n_i + \deg (dx_1)}{2}\\
&=& \frac{2n_i + 2 g_i  - 2}{2}\\
&=& n_i + g_i - 1,
\end{eqnarray*}
where $g_i$ is the genus of $F_i/\mathbf{F}_{q^2}$.

Let $j$ be a nonnegative integer.
Since $\sigma (G_0 + j P_\infty) = G_0 + j P_\infty$,
it satisfies the condition on $G$ in Proposition \ref{prop2}.
Let $H = (P_1 + \cdots + P_n + \sigma P_1 + \cdots + \sigma P_n)
-(G_0 + j P_\infty) + (\eta) = G_0 - jP_\infty$.
Since $G + jP_\infty \geq H$,
$C(G+jP_\infty)^{\perp\mathrm{s}} \subseteq C(G+jP_\infty)$.
By Corollary \ref{coro3} we can construct an $[[n_i, k_{ij},
d_{ij}]]$ quantum stabilizer code with
\[
k_{ij} \geq j,\; d_{ij} \geq (n_i - g_i -j +1)/2.
\]

Let $R$ be a real number such that $0 \leq R\leq 1$,
and set $j$ to $ \lfloor R n_i \rfloor$.
By Theorem \ref{th:ak} we can construct
a sequence of $[[n_i, k_i, d_i]]$ binary quantum
stabilizer codes with
\begin{eqnarray*}
\liminf_{i\rightarrow\infty} k_i/n_i & \geq & R,\\
\liminf_{i\rightarrow\infty} d_i/n_i & \geq & \delta = \frac{1-R-2/(2^m-1)}{4m},
\end{eqnarray*}
because $n_i / g_i$ converges\footnote{%
In the versions 1 and 2 of this eprint,
the author miscalculated $\lim_{i\rightarrow\infty} n_i/g_i$
as $2^m-1$, which led the wrong value of $R^{(1)}_m(\delta)$.
The author apologize for misleading the reader.}
to $(2^m-1)/2$ as $i\rightarrow \infty$ \cite{garcia95}.
Simple calculation shows $R = R_m^{(1)}(\delta)$.
\end{proof}

By choosing an appropriate value $m$ for every $\delta$,
we can construct a sequence of $[[n_i,k_i,d_i]]$ binary
quantum codes
with
\[
\liminf_{i\rightarrow\infty} k_i / n_i \geq R^{(1)}(\delta),\;
\liminf_{i\rightarrow\infty} d_i / n_i \geq \delta,
\]
where
\[
R^{(1)}(\delta) = R_m^{(1)}(\delta)
\mbox{ for } \frac{2^{m-1}}{(2^m-1)(2^{m+1}-1)} \leq \delta
\leq \frac{2^{m-2}}{(2^{m-1}-1)(2^{m}-1)}.
\]
The function $R^{(1)}(\delta)$ is plotted in Figure \ref{fig1}.

\subsection{Improvement of the Ashikhmin-Litsyn-Tsfasman construction}\label{sec32}
In \cite{ashikhmin01}
Ashikhmin et~al.\ constructed an asymptotically good sequence
of binary quantum codes from self-orthogonal classical algebraic
geometry codes.
In their construction they do not use at least $g$ points on the curve
(see Remark below Theorem 4 of \cite{ashikhmin01}),
where $g$ is the genus of the curve.

By \cite[Proposition VII.1.2]{bn:stichtenoth}
we have $C(G_0 + j P_\infty) \supseteq C(G_0 + j P_\infty)^\perp$
for the code $C(G_0 + j P_\infty)$ over $\mathbf{F}_{2^{2m}}$
constructed in the proof
of Proposition \ref{prop:sequence},
where
$C(G_0 + j P_\infty)^\perp$ is the dual code
of $C(G_0 + j P_\infty)$ with respect to
the standard inner product.
In the construction of $C(G_0 + j P_\infty)$
we asymptotically use all the points on the curve.
Therefore we can construct a better sequence of binary quantum codes
if we use $C(G_0 + j P_\infty)$ in the construction of Ashikhmin et~al.

Let us calculate the asymptotic parameters of the sequence
constructed by the method of Ashikhmin et~al.\  with $C(G_0 + j P_\infty)$.
Let $N_i$ ($= 2n_i$) be the code length of $C(G_0 + j P_\infty)$ as a classical
linear code.
We have
\[
\dim C(G_0 + j P_\infty) \geq \dim (G_0 + j P_\infty) \geq j + N_i/2,
\]
and the minimum Hamming distance of $C(G_0 + j P_\infty)$ is not less than
\[
N_i/2 -  g_i + 1 - j.
\]

By the construction of binary quantum codes in \cite{ashikhmin01},
from the inclusion of classical codes
\begin{eqnarray*}
&&
C\left(G_0 + \left(\left(\frac{1}{2}-\delta'\right)N_i - g_i + 1\right)P_\infty
\right)^\perp \\*
&\subset&
C\left(G_0 + \left(\left(\frac{1}{2}-\delta'\right)N_i - g_i + 1\right)P_\infty
\right) \\*
&\subset&
C\left(G_0 + \left(\left\lfloor\left(
\frac{1}{2}-\frac{2}{3}\delta'\right)N_i\right\rfloor - g_i + 1\right)P_\infty
\right)
\end{eqnarray*}
for $0\leq \delta' \leq 1/2- g_i/N_i$,
we can construct
$[[ N_i, k_i, d_i]]$ binary quantum codes with
\[
k_i \geq \left(1-\frac{5}{3}\delta'\right) N_i - 2 g_i + 1,\;
d_i \geq \frac{\delta' N_i}{2m}.
\]
Since $\lim_{i\rightarrow\infty} N_i/g_i = 2^m - 1$ \cite{garcia95},
by setting $\delta = \delta'/2m$
we have
\[
\liminf_{i\rightarrow\infty} \frac{k_i}{N_i} \geq
R_m^{(\mathrm{ALT})} (\delta),\;
\liminf_{i\rightarrow\infty} \frac{d_i}{N_i} \geq \delta,
\]
where 
\begin{equation}
R_m^{(\mathrm{ALT})} (\delta) = 1 - \frac{10}{3}m\delta - \frac{2}{2^m-1}.
\label{improvedalt}
\end{equation}
It is clear that Eq.\ (\ref{improvedalt}) is larger than
Eq.\ (21) of \cite{ashikhmin01}.

By choosing an appropriate value $m$ for every $\delta$,
we can construct a sequence of $[[N_i,k_i,d_i]]$ binary quantum codes
with
\[
\liminf_{i\rightarrow\infty} \frac{k_i}{N_i} \geq R^{(\mathrm{ALT})}(\delta),\;
\liminf_{i\rightarrow\infty} \frac{d_i}{N_i} \geq \delta,
\]
where
$R^{(\mathrm{ALT})}(\delta) = R_m^{(\mathrm{ALT})}(\delta)$ for
\[
\frac{3 \cdot 2^{m}}{5(2^m-1)(2^{m+1}-1)} \leq \delta
\leq 
\min\left\{\frac{5}{84},
\frac{3 \cdot 2^{m-1}}{5(2^{m-1}-1)(2^{m}-1)}\right\}.
\]

\begin{figure}[tb!]
\includegraphics*[width=\textwidth]{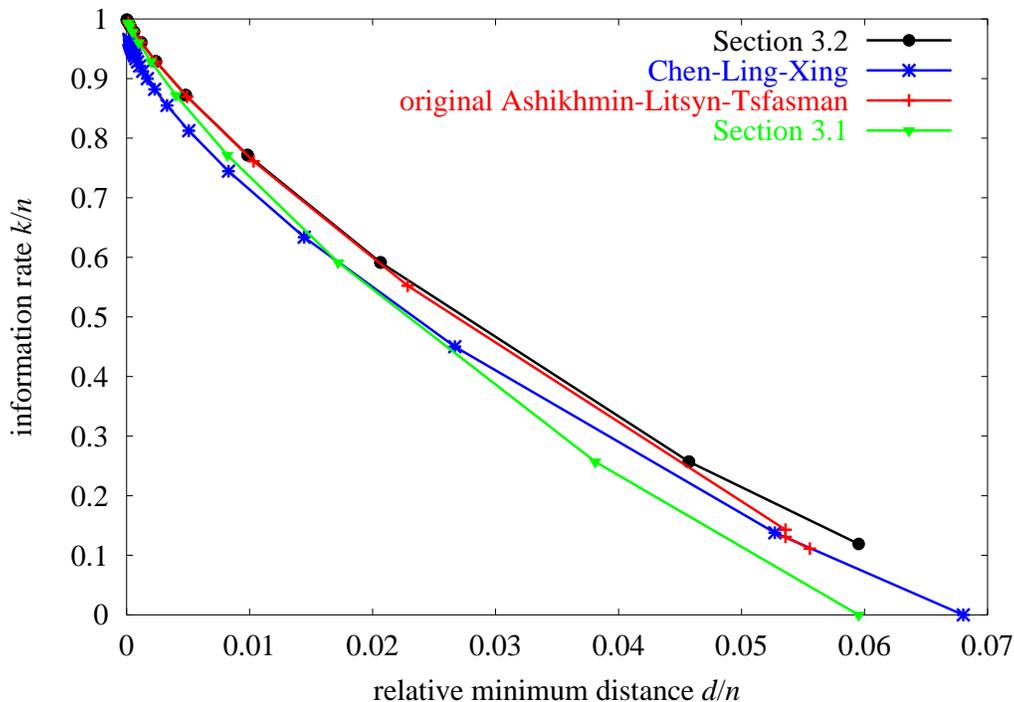}
\caption{Asymptotically good sequences of quantum codes (color)}\label{fig1}
\end{figure}

Ashikhmin et~al.\ \cite{ashikhmin01}
constructed an asymptotically good sequence of binary quantum codes
from algebraic curves.
Chen et~al.\ \cite{chen01b} constructed a sequence
based on the idea in \cite{xing99} better than \cite{ashikhmin01}
in certain range of parameters.
Their sequences and the sequences in this section are compared
in Figure~\ref{fig1}.
Note that Chen \cite{chen01a} also proposed the same construction
of quantum codes as \cite{ashikhmin01}.


\begin{thebibliography}{10}
\small
\setlength{\topsep}{0mm}
\setlength{\partopsep}{0mm}
\setlength{\itemsep}{0mm}
\setlength{\parsep}{0mm}
\setlength{\parskip}{0mm}

\bibitem{ashikhmin00}
A.~Ashikhmin and E.~Knill.
\newblock Nonbinary quantum stabilizer codes.
\newblock May 2000, quant-ph/0005008.

\bibitem{ashikhmin01}
A.~Ashikhmin, S.~Litsyn, and M.~A. Tsfasman.
\newblock Asymptotically good quantum codes.
\newblock {\em Phys.\ Rev.\ A}, 63(3):032311, Mar. 2001, quant-ph/0006061.

\bibitem{bennett01}
C.~H. Bennett, P.~W. Shor, J.~A. Smolin, and A.~V. Thapliyal.
\newblock Entanglement-assisted capacity of a quantum channel and the reverse
  {S}hannon theorem.
\newblock June 2001, quant-ph/0106052.

\bibitem{bennett92}
C.~H. Bennett and S.~J. Wiesner.
\newblock Communication via one- and two-particle operations on
  {E}instein-{P}odolsky-{R}osen states.
\newblock {\em Phys.\ Rev.\ Lett.}, 69(20):2881--2884, Nov. 1992.

\bibitem{calderbank97}
A.~R. Calderbank, E.~M. Rains, P.~W. Shor, and N.~J.~A. Sloane.
\newblock Quantum error correction and orthogonal geometry.
\newblock {\em Phys.\ Rev.\ Lett.}, 78(3):405--408, Jan. 1997,
  quant-ph/9605005.

\bibitem{calderbank98}
A.~R. Calderbank, E.~M. Rains, P.~W. Shor, and N.~J.~A. Sloane.
\newblock Quantum error correction via codes over {GF(4)}.
\newblock {\em IEEE Trans.\ Inform.\ Theory}, 44(4):1369--1387, July 1998,
  quant-ph/9608006.

\bibitem{chen01a}
H.~Chen.
\newblock Some good quantum error-correcting codes from algebraic-geometric
  codes.
\newblock {\em IEEE Trans.\ Inform.\ Theory}, 47(5):2059--2061, July 2001,
  quant-ph/0107102.

\bibitem{chen01b}
H.~Chen, S.~Ling, and C.~Xing.
\newblock Asymptotically good quantum codes exceeding the
  {Ashikhmin-Litsyn-Tsfasman} bound.
\newblock {\em IEEE Trans.\ Inform.\ Theory}, 47(5):2055--2058, July 2001.

\bibitem{farran00}
J.~I. Farr\'an.
\newblock Decoding algebraic geometry codes by a key equation.
\newblock {\em Finite Fields Appl.}, 6(3):207--217, July 2000, math.AG/9910151.

\bibitem{garcia95}
A.~Garcia and H.~Stichtenoth.
\newblock A tower of {A}rtin-{S}chreier extensions of function fields,
  attaining the {D}rinfeld-{V}ladut bound.
\newblock {\em Invent.\ Math.}, 121(1):211--222, July 1995.

\bibitem{gottesman96}
D.~Gottesman.
\newblock Class of quantum error-correcting codes saturating the quantum
  {H}amming bound.
\newblock {\em Phys.\ Rev.\ A}, 54(3):1862--1868, Sept. 1996, quant-ph/9604038.

\bibitem{kak99}
S.~Kak.
\newblock The initialization problem in quantum computing.
\newblock {\em Found.\ Phys.}, 29(2):267--279, Feb. 1999, quant-ph/9811005.

\bibitem{knill96a}
E.~Knill.
\newblock Non-binary unitary error bases and quantum codes.
\newblock Aug. 1996, quant-ph/9608048.

\bibitem{matsumotoerror}
R.~Matsumoto.
\newblock Fidelity of a $t$-error correcting quantum code with more than $t$
  errors.
\newblock {\em Phys.\ Rev.\ A}, 64(2):022314, Aug. 2001, quant-ph/0011047.

\bibitem{peterson}
W.~W. Peterson and E.~J. Weldon, Jr.
\newblock {\em Error-Correcting Codes}.
\newblock MIT Press, Cambridge, Massachusetts, 2nd edition, 1972.

\bibitem{preskill98}
J.~Preskill.
\newblock Lecture Notes for Physics 229: Quantum Information and Computation,
\texttt{http://www.theory.caltech.edu/people/preskill/ph229},
1998.

\bibitem{rains97}
E.~M. Rains.
\newblock Nonbinary quantum codes.
\newblock {\em IEEE Trans.\ Inform.\ Theory}, 45(6):1827--1832, Sept. 1999,
  quant-ph/9703048.

\bibitem{shor95}
P.~W. Shor.
\newblock Scheme for reducing decoherence in quantum computer memory.
\newblock {\em Phys.\ Rev.\ A}, 52(4):2493--2496, Oct. 1995.

\bibitem{steane96b}
A.~M. Steane.
\newblock Error correcting codes in quantum theory.
\newblock {\em Phys.\ Rev.\ Lett.}, 77(5):793--797, July 1996.

\bibitem{stichtenoth88}
H.~Stichtenoth.
\newblock Self-dual {G}oppa codes.
\newblock {\em J.\ Pure Appl.\ Algebra}, 55(1,2):199--211, Nov. 1988.

\bibitem{bn:stichtenoth}
H.~Stichtenoth.
\newblock {\em Algebraic Function Fields and Codes}.
\newblock Springer-Verlag, Berlin, 1993.

\bibitem{xing99}
C.~Xing, H.~Niederreiter, and K.~Y. Lam.
\newblock A generalization of algebraic-geometry codes.
\newblock {\em IEEE Trans.\ Inform.\ Theory}, 45(7):2498--2501, Nov. 1999.

\end{thebibliography}

\end{document}